\documentclass[aps,reprint]{revtex4}
\pdfoutput=1
\usepackage{graphicx,epsf,epsfig}
\usepackage{amsmath,amsfonts,amssymb}
\usepackage{paralist}
\usepackage{accents}
\usepackage{epstopdf}
\usepackage{footnote}
\usepackage{soul}
\usepackage{color}

\newcommand{\be}{\begin{equation}}
\newcommand{\ee}{\end{equation}}

\newcommand{\ba}{\begin{array}{c}}
\newcommand{\baz}{\begin{array}{cc}}
\newcommand{\bad}{\begin{array}{ccc}}
\newcommand{\bav}{\begin{array}{cccc}}
\newcommand{\baf}{\begin{array}{ccccc}}
\newcommand{\bag}{\begin{array}{cccccc}}
\newcommand{\ea}{\end{array}}

\begin{document}

\title{$A_4$ Flavor Symmetry Model for Dirac-Neutrinos and Sizable $U_{e3}$}
\author{Nina Memenga}
\email{nina.memenga@gmx.de}
\author{Werner Rodejohann}
\email{werner.rodejohann@mpi-hd.mpg.de}
\affiliation{Max-Planck-Institut f\"ur Kernphysik, Saupfercheckweg 1, 69117 Heidelberg, Germany}
\author{He Zhang}
\email{he.zhang@mpi-hd.mpg.de}
\affiliation{Max-Planck-Institut f\"ur Kernphysik, Saupfercheckweg 1, 69117 Heidelberg, Germany}

\begin{abstract}
Models based on flavor symmetries are the most often studied approaches
to explain the unexpected
structure of lepton mixing. In many flavor symmetry groups a product
of two triplet representations contains a symmetric and an
anti-symmetric contraction to a triplet. If this product of two triplets corresponds to
a Majorana mass term, then the anti-symmetric part vanishes, and in
economic models tri-bimaximal mixing is achieved. If neutrinos are
Dirac particles, the anti-symmetric part is however present and
leads to deviations from tri-bimaximal mixing, in particular
non-zero $U_{e3}$. Thus, the non-vanishing value of $U_{e3}$ and the
nature of the neutrino are connected. We illustrate this with a
model based on $A_4$ within the framework of a neutrinophilic 2
Higgs doublet scenario.

\end{abstract}
\pacs{14.60.Pq; 11.30.Hv, 14.60.St}

\maketitle

\section{\label{sec:intro}Introduction}
Fermions of the Standard Model (SM) have an interesting property: they mix among each other.
While describing mixing is straightforward,
explaining the observed values is not possible in the SM.
Determining the theory behind fermion mixing is therefore one of the most pressing
issues in current particle physics.

In particular the unlikely and peculiar mixing structure of the leptons, that has been
determined experimentally in the last two decades, made the situation more puzzling. For
quite some time the so-called tri-bimaximal mixing (TBM) scheme was considered to be an excellent
description of the leptonic mixing matrix \cite{tbm}:
\be \label{eq:tbm}
U_{\rm TBM} = \left(
\bad
\sqrt{\frac 23} & \sqrt{\frac 13}  & 0 \\
-\sqrt{\frac 16}  & \sqrt{\frac 13}  & -\sqrt{\frac 12} \\
-\sqrt{\frac 16}  & \sqrt{\frac 13}  & \sqrt{\frac 12}
\ea
\right) .
\ee
Apparently, some symmetry input is required to generate such a mixing pattern. These flavor
symmetries assume that the left- and right-handed leptons, as well as new particles, transform
as irreducible representations of some (typically discrete) symmetry group.  In the vast majority
of theoretical approaches to Eq.\ (\ref{eq:tbm}), the rotation group of the tetrahedron, $A_4$, is
applied. This symmetry was first proposed in Ref.\ \cite{Ma:2001dn}. 
Surprisingly, it was possible to construct rather economic and straightforward models
\cite{revs}.
The group $A_4$ is a natural choice for the flavor symmetry, since it is the smallest group with
a three-dimensional representation $\bf 3$.
Thus, the three generations of left-handed weak lepton doublets could be unified and identified
with the $\bf 3$.
Furthermore, $A_4$ has three one-dimensional irreducible representations
$\bf 1$, $\bf 1'$ and $\bf 1''$, which can be identified with the three right-handed charged
lepton singlets. We note that to the best of our
knowledge, all of the literally hundreds of flavor symmetry models,
be it with $A_4$ or any other group, were exclusively assuming Majorana neutrinos (see for instance
Ref.~\cite{Barry:2010zk} for a list and classification of $A_4$ models).
This assumption can only be tested in experiments looking for neutrinoless double beta decay,
$(A,Z) \to (A,Z+2) + 2 e^-$, and currently various collaborations are performing searches
for this process \cite{Rodejohann:2011mu}. The question on whether neutronis are Dirac or
Majorana particles is one of the most interesting and important ones of the field.
In this paper, we will assume neutrinos to be Dirac particles. Hence, in the
absence of any other lepton number violating physics, there will be no neutrinoless double
beta decay.

A recent observation, that cast some doubt on tri-bimaximal mixing
was the largish value of the mixing matrix element $|U_{e3}| \simeq
0.16$, that has been determined by reactor neutrino experiments
\cite{t13}. Also the best-fit
points~\cite{Fogli:2012ua,Tortola:2012te,GonzalezGarcia:2012sz}
of $U_{e2}$ and $U_{\mu 3}$
deviate, though less strongly, from the predictions of
Eq.~(\ref{eq:tbm}). Generating sizable corrections to the mixing
scheme is possible, but requires for instance large higher
dimensional contributions to the models, or large neutrino masses
and/or $\tan \beta$ in renormalization group effects. Here care has
to be taken to guarantee that $\delta U_{e3}
> \delta U_{e2, \mu 2}$. Of course, it is also possible to construct
models that give mixing schemes different from TBM, in particular
with initially non-zero $U_{e3}$. However, those models are typically less
economic than the ones leading to Eq.~(\ref{eq:tbm}), as they
usually involve much larger groups, see
for instance Ref.\ \cite{Holthausen:2012wt}. In this work we wish to
keep the minimality of typical $A_4$ models and introduce a new way
to generate non-zero $U_{e3}$ and other deviations from TBM.

A typical ingredient in flavor symmetry models that use a group
involving irreducible triplet representations, is that there will be
a coupling of three triplets, ${\bf 3} \times {\bf 3} \times {\bf
3}$. Here two of the triplets are left- and/or right-handed
fermions, and the third triplet is a set of scalar flavon fields or
Higgs doublets. In many phenomenologically interesting
flavor symmetries (e.g.\ $A_4$, $T'$, $\Delta(3n^2)$ and
$\Delta(6n^2)$~\cite{Luhn:2007uq}) the tensor product
of two triplets contains both symmetric and anti-symmetric
contractions to a triplet: 
\be \label{eq:33=sa} 
{\bf 3} \times {\bf 3} = {\bf
3}_{\rm s} + {\bf 3}_{\rm a} + \ldots 
\ee 
Here ${\bf 3}_{\rm s}$
and ${\bf 3}_{\rm a}$ are a symmetric and an anti-symmetric
combination of the components in the two multiplied triplets
\footnote{Actually, the fact that the combinations are
(anti-)symmetric combinations of the individual triplet components
depends on the basis for the representations of the group. Our
analysis is not affected by the basis choice, but becomes clearer in
a basis in which the triplet product contains a symmetric and an
anti-symmetric combination.}. The remaining terms in
Eq.~(\ref{eq:33=sa}) depend on the group, and in $A_4$ are given by
${\bf 1} + {\bf 1'}+ {\bf 1''} $. If the mass term that is
constructed from Eq.~(\ref{eq:33=sa}) is a Majorana mass term, i.e.,
the two triplets are left- or right-handed neutrinos, then the
anti-symmetric combination vanishes. In these theories, TBM is then
eventually achieved.

Our observation is the following: if neutrinos are Dirac instead of
Majorana particles, and their mass term depends on a product of two triplets of fermions, where
one is left- and the other right-handed,
then the anti-symmetric ${\bf 3}_{\rm a}$ term
will not vanish in general. Hence, with essentially the same
particle content and transformation properties under the flavor
symmetry group, deviations from tri-bimaximal mixing
will arise, in particular non-zero $U_{e3}$.
Thus, the nature of the neutrino and the non-vanishing value of
$U_{e3}$ are linked. This provides a new way to accommodate non-zero
$U_{e3}$ in flavor symmetry models, while keeping the economic
structure of typical models.

We note that the property given in Eq.~(\ref{eq:33=sa}) does not
hold solely for $A_4$, but also for other popular groups. Moreover, the idea we propose can
also be applied to any mixing scheme other than TBM. The example that we will give to
illustrate our observation will for definiteness be in the framework
of $A_4$ and tri-bimaximal mixing.

We also need to specify the mechanism that guarantees the Dirac
nature of the neutrinos. Usually the flavor symmetry models
available in the literature assume Majorana neutrinos, and generate
Majorana masses either via an effective operator, or within the type
I or II seesaw mechanism. Our choice is the ``neutrinophilic'' 2
Higgs Doublet Model for Dirac neutrinos as considered in
\cite{Wang:2006jy,Gabriel:2006ns,Davidson:2009ha},
in which a second Higgs doublet is introduced which
exclusively couples to neutrinos. The smallness of their masses is
explained by the small vacuum expectation value of this doublet. A consistent 
2 Higgs doublet framework that incorporates such a small vacuum expectation value 
requires soft breaking of the underlying symmetry by a bilinear 
term that couples the two Higgs doublet. This general idea was 
first introduced in Ref.\ \cite{Ma:2000cc}. 
We note that this type of model is not in conflict with the recently
observed \cite{lhc} new particle at the Large Hadron Collider, and
could in fact, as any 2 Higgs doublet model \cite{Branco:2011iw}, be
used to explain an excessive decay rate of that particle into two
photons (if the rather mild preference for this remains with more
data). In the next Sections we will first demonstrate how a typical
$A_4$ model for Majorana neutrinos works, before modifying it to the
Dirac neutrino case, realized in the framework of the
neutrinophilic 2 Higgs Doublet Model.

\section{\label{sec:model}Majorana model example}

In order to illustrate the situation in an easier manner, we start
with a brief review of an economic and minimal $A_4$ ``role model'',
in which the full symmetry group is given by $G_{\rm SM} \otimes A_4
\otimes Z_3$ ($G_{\rm SM}$ being the Standard Model
gauge group), and neutrinos are assumed to be
Majorana particles. Here an additional cyclic $Z_3$ is added in
order to disentangle the flavons for the charged lepton and neutrino
sectors. Similar to the model addressed in
Ref.~\cite{Altarelli:2005yx}, apart from the SM particle content we introduce three
right-handed neutrinos assigned to the three-dimensional
representation of $A_4$, together with three sets of flavon fields
$\varphi$, $\varphi'$ and $\xi$ (see Table~\ref{tab:1} for details
of the particle assignments).
\begin{table}[t]
\caption{\label{tab:1}Particle assignments for the Majorana neutrino
model. The additional $Z_3$ symmetry decouples the charged lepton and neutrino
sectors, and $\omega=e^{i2\pi/3}$ is the complex cube-root of
unity.}
\begin{center}
\begin{tabular}{c|cccc|cccc|c}
\hline \hline
Field & $\ell$ & $e_R$ & $\mu_R$ & $\tau_R$ & $H$ & $\varphi$ &
$\varphi'$ & $\xi$ & $\nu_R$ \\ \hline \hline
$A_4$ & $\bf 3$ & $\bf 1$ & $\bf 1''$ & $\bf 1'$ & $\bf 1$ & $\bf 3$ & $\bf 3$ & $\bf 1$  & $\bf 3$ \\
\hline
$Z_3$ & $\omega$ & $\omega^2$ & $\omega^2$ & $\omega^2$ & $1$ & $1$ & $\omega^2$ & $\omega^2$ & $\omega^2$  \\
\hline \hline
\end{tabular}
\end{center}
\end{table}
The invariant Lagrangian at leading order
can be written as
\begin{eqnarray}\label{eq:Lang-M} 
{\cal L} & = & \frac{y_e}{\Lambda}  \left( \varphi\overline{\ell}
\right)_{1} H e_R + \frac{y_\mu}{\Lambda} \left(
\varphi\overline{\ell} \right)_{1'} H \mu_R + \frac{y_\tau}{\Lambda}
\left( \varphi\overline{\ell} \right)_{1''}
H \tau_R 
+ y_D \left( \overline{\ell}
 \nu_R \right)_1 \tilde{H}  + x_A \xi \left( \overline{\nu^c_R} \nu_R \right)_1+ x_B \varphi' \left( \overline{\nu^c_R} \nu_R\right)_{3_{\rm s}} 
\; ,
\end{eqnarray}
where $\tilde H\equiv i\tau_2 H^*$, and the well-known tensor product rules of
$A_4$ can be read from e.g.~Ref.~\cite{He:2006dk}, and are given for completeness
also in our appendix.
The notation in Eq.~(\ref{eq:Lang-M}) is as usual, $\left(
\varphi\overline{\ell} \right)_{1'}$ denotes the part of the triplet product between
$\varphi$ and $\overline{\ell}$ that transforms as $\bf 1'$.
Following Ref.~\cite{Altarelli:2005yx} we assume that the flavon
fields develop a vacuum expectation value (VEV) along the directions
\begin{eqnarray} \label{eq:VEV}
\langle \varphi \rangle = \begin{pmatrix} v & v &
v\end{pmatrix}^T  , ~~\nonumber 
\langle \varphi' \rangle = \begin{pmatrix} 0 & v' & 0\end{pmatrix}^T
,~~
\langle \xi \rangle  =  u \; . \nonumber
\end{eqnarray}
See e.g.~Ref.~\cite{Altarelli:2005yx} for the techniques to achieve this VEV alignment
in a natural way.
We assume in what follows that the usual mechanisms to guarantee such alignment are at work.
After the breaking of the flavor and the electroweak symmetries, the
charged leptons develop a mass term 
\begin{eqnarray}
\frac{\langle H \rangle}{\Lambda}
\left(\overline{e_L},\overline{\mu_L},\overline{\tau_L}\right)
\begin{pmatrix} y_e v & y_\mu v & y_\tau v \cr y_e v & \omega y_\mu v &
\omega^2 y_\tau v \cr y_e v & \omega^2 y_\mu v & \omega y_\tau v
\end{pmatrix} \begin{pmatrix} e_R \cr \mu_R \cr \tau_R \end{pmatrix}
,
\end{eqnarray}
and can be diagonalized by using a bi-unitary
transformation $V^\dagger_L M_\ell V_R = {\rm diag}
(m_e,m_\mu,m_\tau)$. Here $m_f=\sqrt{3} y_f \langle H \rangle
v/\Lambda$ (for $f=e,\mu,\tau$) are the charged-lepton masses and
\begin{eqnarray}
V_L=  \frac{1}{\sqrt{3}}
\begin{pmatrix} 1 & 1 & 1 \cr 1 & \omega & \omega^2 \cr 1 & \omega^2
& \omega
\end{pmatrix} .
\end{eqnarray}
The Dirac neutrino mass matrix is
simply proportional to the unit matrix, $M_D =  y_D \langle
H \rangle {\rm diag} \left(1,1,1\right)$. The right-handed neutrino
mass matrix is found to be
\begin{eqnarray} \label{eq:MR}
M_R = \begin{pmatrix} x_A u & 0 & x_B v'   \cr 0 & x_A u & 0 \cr x_B
v' & 0 & x_A u
\end{pmatrix}  .
\end{eqnarray}
Finally, the mass matrix for the light neutrinos is obtained by
using the standard seesaw formula $M_\nu = M_D M^{-1}_R M^T_D$,
leading to
\begin{eqnarray}\label{eq:mnuM}
M_\nu  = \frac{y^2_D \langle H \rangle^2}{x^2_A
u^2 - x^2_B v'^2}
\begin{pmatrix} x_A u  & 0 & -x_B v'   \cr 0 & \frac{x^2_A
u^2 - x^2_B v'^2}{x_A u} & 0 \cr -x_B v' & 0 & x_A u
\end{pmatrix}  ,
\end{eqnarray}
where we omit the minus sign for simplicity. $M_\nu$ is easily
diagonalized by 
\begin{eqnarray}
V_\nu = \frac{1}{\sqrt{2}}
\begin{pmatrix} 1 & 0 & -1 \cr 0 & \sqrt{2} & 0 \cr 1 & 0 & 1
\end{pmatrix} ,
\end{eqnarray}
and the leptonic flavor mixing matrix stems from the mismatch between
$V_L$ and $V_\nu$,
\begin{eqnarray}
U=V_L^\dagger V_\nu = \begin{pmatrix} \sqrt{\frac{2}{3}} &
\sqrt{\frac{1}{3}} & 0 \cr \frac{1+\omega }{\sqrt{6}} &
\frac{\omega^2}{\sqrt{3}} & \frac{\omega-1}{\sqrt{6}} \cr
\frac{1+\omega^2 }{\sqrt{6}} & \frac{\omega}{\sqrt{3}} &
\frac{\omega^2-1}{\sqrt{6}}
\end{pmatrix} .
\end{eqnarray}
Thus, TBM is obtained, up to irrelevant phases.

We stress here that in Eq.~(\ref{eq:mnuM}) the terms proportional to $x_B$ stem from the
triple-triplet product $\nu_R \times \nu_R \times \varphi' $, where due to the
Majorana nature of the $\nu_R$ only the symmetric contribution of the $\nu_R \times \nu_R$
tensor product survives.

\section{$A_4$ symmetry in the $\nu$2HDM}

In the Dirac neutrino case, we work in the $\nu$2HDM, or
neutrinophilic 2 Higgs Doublet Model. Here, an additional $SU(2)$
doublet $H_\nu$---with the same quantum numbers as the SM Higgs
doublet $H$---is introduced. The flavon content is the same as in
the previous Majorana neutrino model, except that the $Z_3$ charges
of $\varphi$, $\xi$ and the right-handed neutrinos are modified. The
economic structure of minimal $A_4$ models is therefore preserved.
Furthermore, a global $U(1)$ symmetry under which the new Higgs
doublet $H_\nu$ and $\nu_R$ carry charge $+1$ while all the other SM
fields are uncharged is imposed. This $U(1)$ symmetry is needed to
forbid Majorana mass terms for the right-handed neutrinos, and
enforces a Yukawa coupling structure in which only $H_\nu$ couples
to right-handed neutrinos.  
We refer the readers to
Ref.~\cite{Davidson:2009ha} for a detailed description of the
constraints and phenomenology of the $\nu$2HDM. It is enough for our
purposes to know that a consistent and allowed framework is
possible, in which the VEV of the doublet responsible for neutrino
Dirac masses $v_\nu = {\cal O}(\rm eV)$ is small. Therefore, the
smallness of neutrino Dirac masses is explained by the small VEV, and not
by tiny Yukawa couplings.

The particle content and their assignments under $A_4$, $Z_3$ and $U(1)$ are summarized in
Table~\ref{tab:2}.
\begin{table}[t]
\caption{\label{tab:2}Particle assignments of the Dirac flavor $A_4$
model.}
\begin{center}
\begin{tabular}{c|cccc|ccccc|c}
\hline \hline
Field & $\ell$ & $e_R$ & $\mu_R$ & $\tau_R$ & $H$ & $H_\nu$ &
$\varphi$ & $\varphi'$ & $\xi$ & $\nu_R$ \\ \hline \hline
$A_4$ & $\bf 3$ & $\bf 1$ & $\bf 1''$ & $\bf 1'$ & $\bf 1$ & $\bf 1$ &
$\bf 3$ & $\bf 3$ & $\bf 1$  & $\bf 3$ \\
\hline
$U(1)$ & $0$ & $0$ & $0$ & $0$ & $0$ & $1$ & $0$ & $0$ & $0$ & $1$  \\
\hline
$Z_3$ & $\omega$ & $\omega^2$ & $\omega^2$ & $\omega^2$ & $1$ & $1$ & $1$ & $\omega$ & $\omega$ & $\omega$  \\
\hline \hline
\end{tabular}
\end{center}
\end{table}
The invariant Lagrangian now reads
\begin{eqnarray} \label{eq:Lang}
\hspace{-.6cm}{\cal L} & = & \frac{y_e}{\Lambda}  \left( \varphi\overline{\ell}
\right)_{1} H e_R + \frac{y_\mu}{\Lambda} \left(
\varphi\overline{\ell} \right)_{1'} H \mu_R + \frac{y_\tau}{\Lambda}
\left( \varphi\overline{\ell} \right)_{1''} H \tau_R
+ \frac{y_s}{\Lambda}  \left( \varphi'\overline{\ell}
 \right)_{3_{\rm s}} \tilde{H}_\nu \nu_R  + \frac{y_a}{\Lambda}  \left( \varphi'\overline{\ell}
 \right)_{3_{\rm a}} \tilde{H}_\nu \nu_R +  \frac{y_x}{\Lambda}  \left(
 \overline{\ell} \tilde{H}_\nu \nu_R
 \right)_{1}  \xi \,.
\end{eqnarray}
We should note that in the general neutrinophilic $\nu$2HDM 
the small VEV $v_\nu$ is generated  by 
introducing an explicit and soft $U(1)$ breaking term $m_{12}^2 \, H^\dagger \, H_\nu$ 
in the Higgs potential \cite{Davidson:2009ha} 
(first proposed in a Majorana neutrino model in Ref.\ \cite{Ma:2000cc}). 
By attributing $U(1)$ quantum numbers to $H_\nu$
and $\nu_R$, and by breaking it softly only by the $m_{12}^2 \,
H^\dagger \, H_\nu$ term, one has actually imposed a residual symmetry in the Lagrangian, namely
$U(1)$ lepton number, thus avoiding a Majorana mass term.


It is important to stress that, since neutrinos are Dirac
particles, the neutrino mass term is from the triple-triplet product
$\varphi' \times  \overline{\ell} \times \nu_R $.
With the property in Eq.~(\ref{eq:33=sa}) it is clear that
the triplet product of $ \overline{\ell}$ and $\nu_R $ contains a
symmetric and an anti-symmetric part. For Majorana neutrinos, cf.~Eq.~(\ref{eq:Lang-M}),
the mass term would depend on
$\varphi' \times  \nu_R \times \nu_R $, not containing an anti-symmetric term.
As we will see, the anti-symmetric part of the Dirac mass term in Eq.~(\ref{eq:Lang})
is, in essentially the same model as the one leading
to Eq.~\eqref{eq:Lang-M}, responsible for deviations from TBM, in particular non-zero $U_{e3}$.

Taking the same VEV alignments as before, the charged lepton sector is identical to the
previous model. The neutrino mass matrix is given by
\begin{eqnarray} \label{eq:mnu}
M_\nu = \frac{v_\nu}{\Lambda}
\begin{pmatrix} y_x u & 0 & (y_s+y_a) v'   \cr 0 & y_x u & 0
\cr (y_s-y_a)v' & 0 & y_x u
\end{pmatrix} ,
\end{eqnarray}
where $v_\nu = \langle H_\nu \rangle$ is the VEV of the Higgs doublet responsible
for Dirac neutrino masses.
Note that $M_\nu$ is not symmetric. In particular, the terms proportional to $y_a$ stem from the
anti-symmetric part of the product of the two $A_4$ triplets $\ell$ and $\nu_R$.
Furthermore, the matrix elements in $M_\nu$ are in general complex.
One can, however, take $y_x$ to be real
without loss of generality. 
The physically relevant part of the neutrino mass matrix can be
expressed as the Hermitian matrix ${\cal H} = M_\nu M^\dagger_\nu$,
satisfying the relation $V^\dagger_\nu {\cal H} V_\nu = {\rm diag}
(m^2_1,m^2_2,m^2_3)$, with $m_i$ being the neutrino masses.
Explicitly, one has
\begin{eqnarray} \label{eq:H}
{\cal H} = \begin{pmatrix} |a|^2 + b^2 & 0 & b(a+c^*) \cr 0 & b^2 &
0 \cr b(a^*+c) & 0 & |c|^2 + b^2
\end{pmatrix} ,
\end{eqnarray}
where $a=(y_s+ y_a) v' v_\nu /\Lambda$, $b=y_x u
v_\nu /\Lambda$, $c=(y_s - y_a) v' v_\nu/\Lambda$.
It can be diagonalized by 
\begin{eqnarray}
R_{13}(\theta,\phi)= \begin{pmatrix} \cos\theta & 0 & \sin\theta e^{-i
\phi} \cr 0 & 1 & 0 \cr -\sin\theta e^{i \phi} & 0 & \cos\theta
\end{pmatrix}  ,
\end{eqnarray}
where the rotation angle and phase are given by
\begin{eqnarray}
\sin2\theta = \frac{2|b(a+c^*)|}{m^2_0} \; ,~~ \nonumber 
\tan\phi = \frac{{\rm Im} (a+c^*)}{{\rm Re} (a+c^*)} \; ,
\end{eqnarray}
with
$m^2_0 =\sqrt{(|a|^2-|c|^2)^2 + 4b^2|a+c^*|^2}$.
The neutrino masses are
\begin{eqnarray}
m^2_1 & = & \frac{1}{2}(|a|^2 +|c|^2) + b^2 -\frac{1}{2} m^2_0 \; ,\nonumber \\
m^2_2 & = & b^2 \; , \\
m^2_3 & = & \frac{1}{2}(|a|^2 +|c|^2) + b^2 + \frac{1}{2}m^2_0 \; .
\nonumber
\end{eqnarray}
An interesting relation can be inferred, namely $\Delta
m^2_{23}+\Delta m^2_{21} = -(|a|^2+|c|^2) <0$. An immediately
consequence is that $m_2$ cannot be larger than $m_3$, and the
inverted neutrino mass ordering is therefore not allowed in the
model. The leptonic flavor mixing matrix is
given by $U= V_L^\dagger R_{13}\left(\theta,\phi\right)$, and reads
\begin{eqnarray}
U= \begin{pmatrix} \displaystyle \frac{c_\theta- s_\theta e^{i \phi}
}{\sqrt{3}} &  \displaystyle  \frac{1}{\sqrt{3}} & \displaystyle
\frac{c_\theta + s_\theta e^{-i \phi}}{\sqrt{3}}  \cr \displaystyle
\frac{c_\theta - s_\theta e^{i \phi} \omega}{\sqrt{3}} &
\displaystyle \frac{\omega^2}{\sqrt{3}} &  \displaystyle
\frac{s_\theta e^{-i \phi} + c_\theta \omega}{\sqrt{3}} \cr
\displaystyle \frac{c_\theta - s_\theta e^{i \phi}
\omega^2}{\sqrt{3}} & \displaystyle \frac{\omega}{\sqrt{3}} &
\displaystyle \frac{s_\theta e^{-i \phi} + c_\theta
\omega^2}{\sqrt{3}}
\end{pmatrix} .
\end{eqnarray}
It contains only two real parameters.
\begin{figure*}[t]
\begin{center}
\includegraphics[width=.5\textwidth,height=7cm]{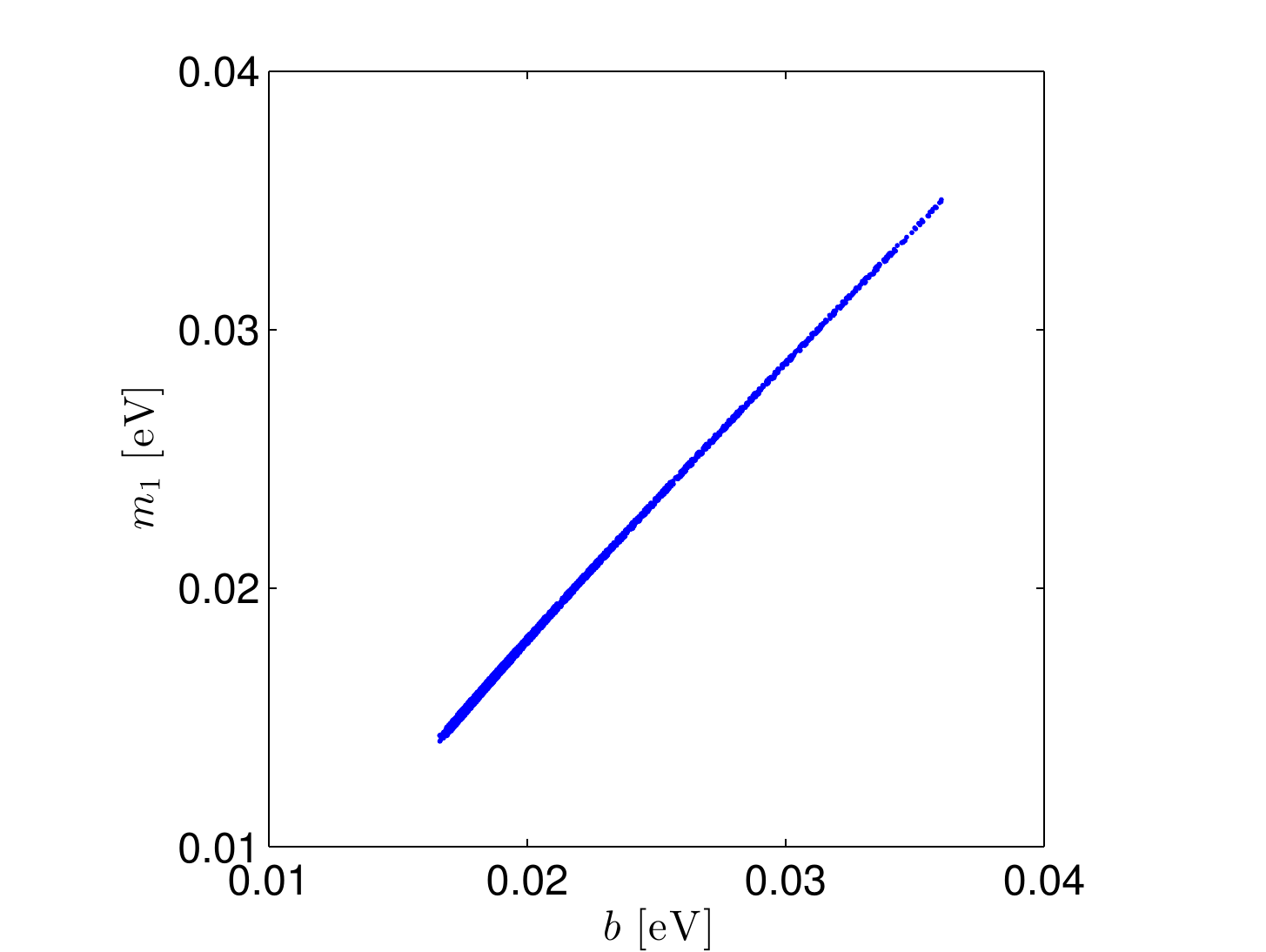}
\hspace{-.3cm}\includegraphics[width=.5\textwidth,height=7cm]{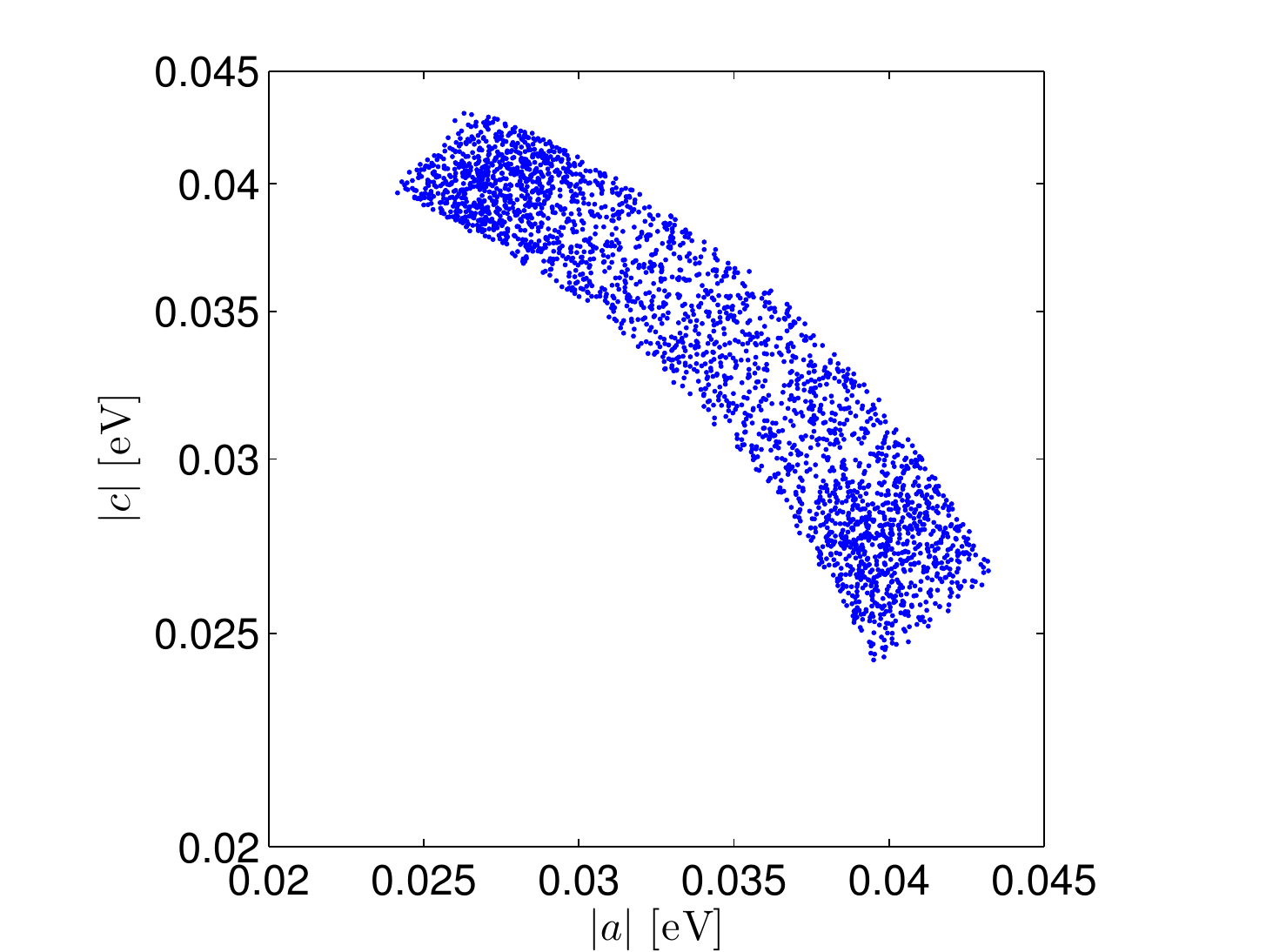}

\includegraphics[width=.5\textwidth,height=7cm]{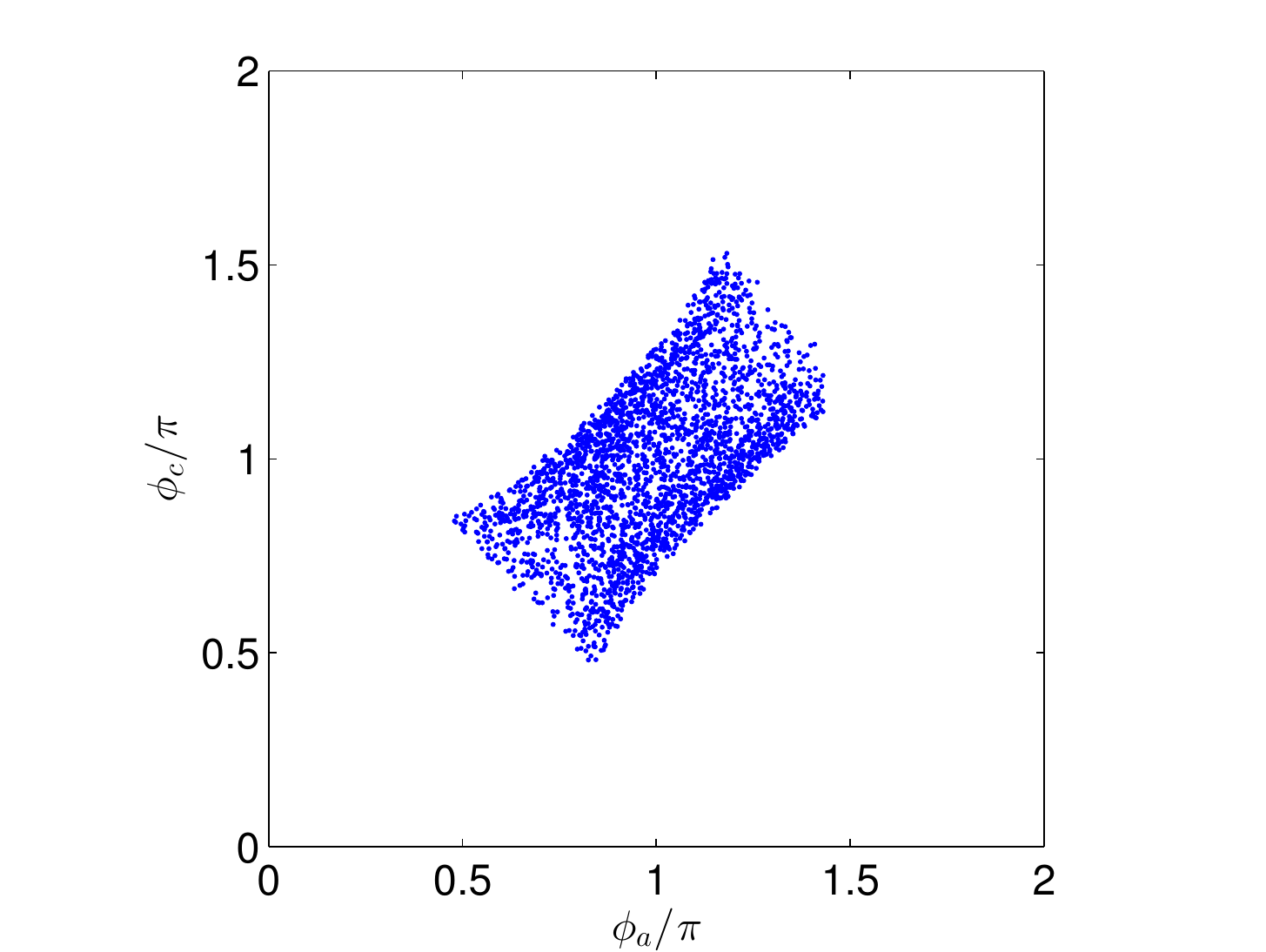}
\hspace{-.3cm}\includegraphics[width=.5\textwidth,height=7cm]{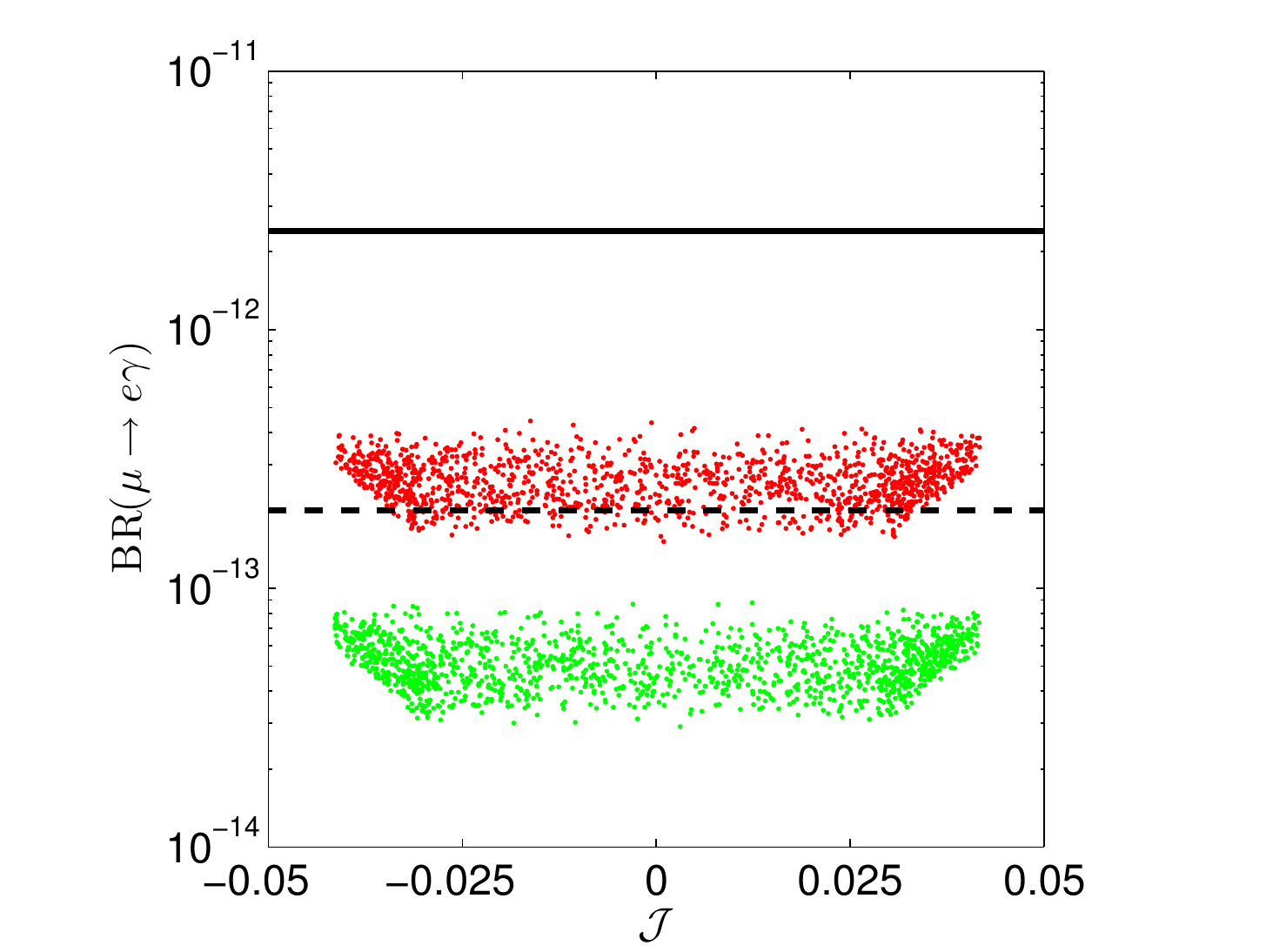}
\caption{ \label{fig:param}Parameter values in Eq.~(\ref{eq:H})
that reproduce the allowed $3\sigma$ ranges of the neutrino parameters.
The right plot in the lower row gives the correlation between leptonic CP violation and
the decay $\mu \to e \gamma$. }
\end{center}
\end{figure*}
This is to be compared to
three experimentally measured neutrino mixing angles and one Dirac
CP phase. The anti-symmetric contribution proportional to $y_a$ in
the Dirac mass matrix (\ref{eq:mnu}) is crucial for deviations from TBM:
in the limit $y_a=0$, which implies $a=c$, it follows that
$c_\theta + s_\theta e^{-i \phi}=0$, and tri-bimaximal mixing is
reproduced. Therefore, the anti-symmetric entry $y_a$ in the Dirac neutrino mass matrix,
that has its origin in the Dirac nature of the neutrinos, causes the deviations from
tri-bimaximal mixing, thereby linking the nature of the neutrino with non-vanishing
$U_{e3}$.

Compared to the exact TBM mixing pattern, the absolute values of the second
column of $U$ remain to be $(\sqrt{1/3},\sqrt{1/3},\sqrt{1/3})^T$,
which leads to the following well-known relations \cite{tm}
\begin{eqnarray} 
\sin^2\theta_{12} =  \frac{1}{3}\frac{1}{1-|U_{e3}|^2} ~, ~~
\cos\delta \tan2\theta_{23} =
\frac{1-2|U_{e3}|^2}{|U_{e3}|\sqrt{2-3|U_{e3}|^2}} \; .
\end{eqnarray}
In our case the Jarlskog invariant is
\begin{eqnarray}
{\cal J}= {\rm Im}\{U_{e1} U_{\mu 2} U^\ast_{e2} U^\ast_{\mu 1} \}
= -\frac{1}{6\sqrt{3}} \cos2\theta \; .
\end{eqnarray}
It is worthwhile to notice that $\cal J$ is independent of the phase
$\phi$.

We proceed with numerical illustrations of the Dirac neutrino model.
Scanning values of the parameters $a$, $b$, $c$ and comparing them
to the $3\sigma$ ranges of the oscillation parameters from
\cite{Tortola:2012te}, we obtain the plots in
Fig.~\ref{fig:param}. Here $\phi_a$ and $\phi_c$ are the phases of $a$ and $c$.
One reads from the plots that the neutrino mass spectrum tends to be
hierarchical, i.e.~$|b|=m_2 \sim [0.015,0.035]~{\rm eV} $.
Since the charged component of the second Higgs doublet
mediates lepton flavor violating processes, we also show the
branching ratio of $\mu \rightarrow e \gamma$,
\begin{equation}
{\rm BR}(\mu \to e \gamma) = \frac{\alpha}{96 \pi} \frac{|{\cal H}_{e\mu}|^2}{8 \, G_F^2 \, M_{H^+}^4 \,
v_\nu^4} \, ,
\end{equation}
versus the Jarlskog invariant $\cal J$, which is proportional to the imaginary part of
${\cal H}_{e\mu} \, {\cal H}_{\mu\tau} \, {\cal H}_{\tau e}$.
We have taken $M_{H^+}=100~{\rm GeV}$ (red, upper points)
and $M_{H^+}=150~{\rm GeV}$ (green, lower points) as two examples,
together with $v_\nu=4~{\rm eV}$. The current upper bound on
the branching ratio, ${\rm BR} (\mu
\rightarrow e\gamma)<2.4\times 10^{-12}$ at $90\%~{\rm C.L.}$ \cite{meg}, is
also indicated on the plot using a black line, a possible future limit of
$2 \times 10^{-13}$ is also indicated. We have nothing to add to the study of the usual
Higgs phenomenology of the $\nu$2HDM \cite{Davidson:2009ha}, the decay $\mu \to e \gamma$
is the only interesting place where some non-trivial correlation exists.


\begin{figure*}[t]
\begin{center}
\includegraphics[width=.5\textwidth]{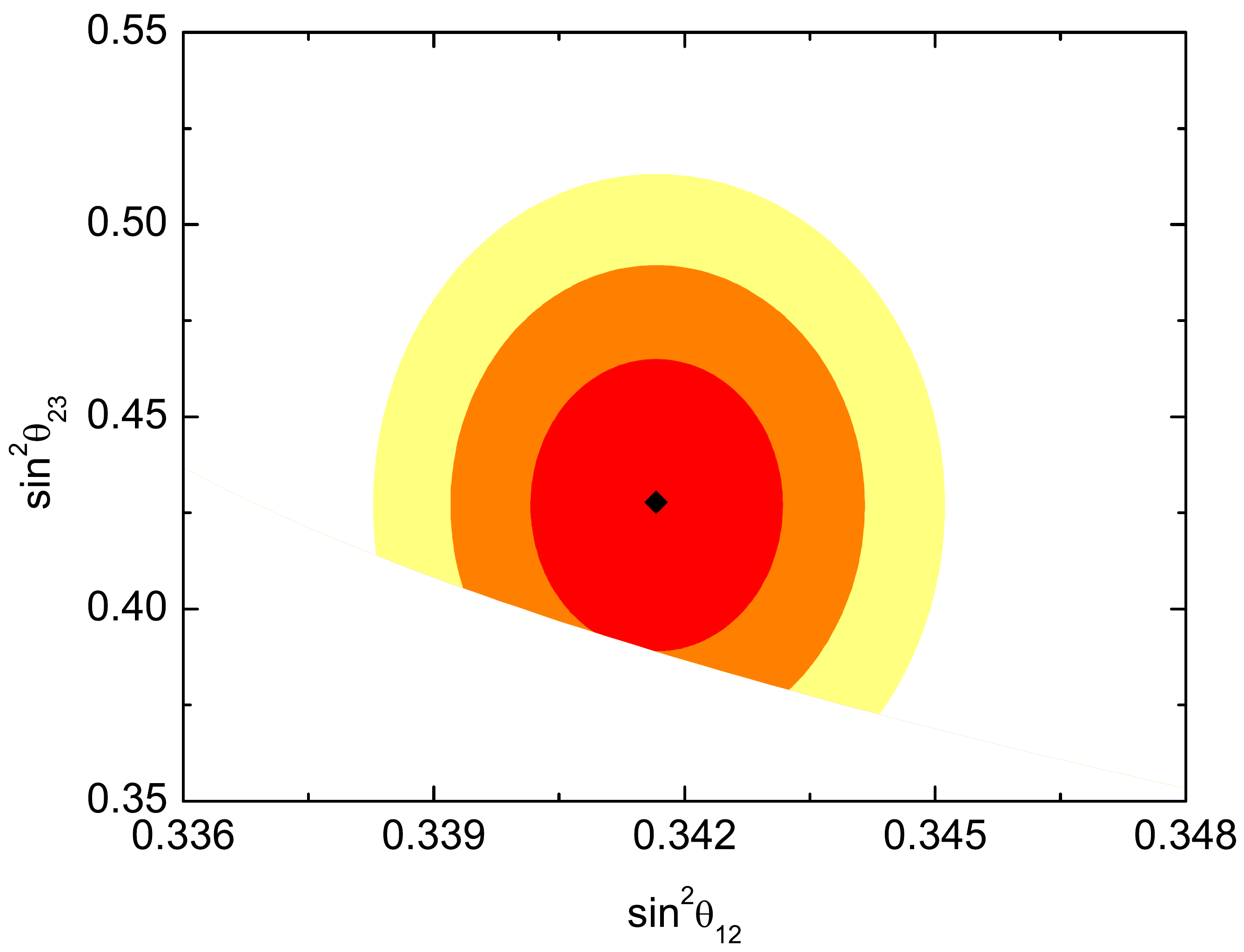}
\hspace{-.3cm}\includegraphics[width=.5\textwidth]{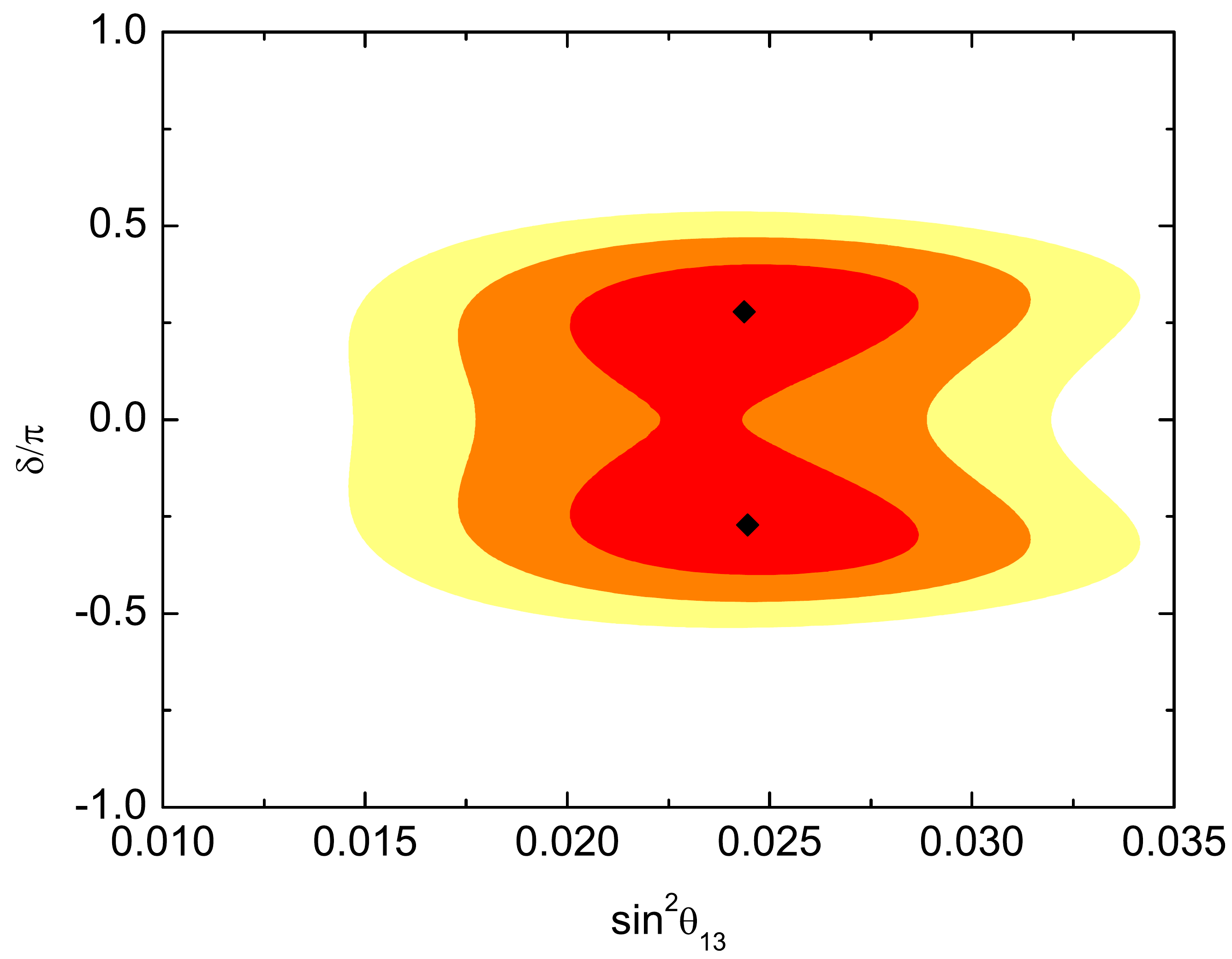}
\caption{ \label{fig:contour}
Allowed region of the physical observables at 1$\sigma$, 2$\sigma$, and 3$\sigma$ C.L.}
\end{center}
\end{figure*}

In order to demonstrate the predictive power of our model for the
neutrino mixing parameters, we compare (following the strategy of Ref.\
\cite{Rodejohann:2012cf}) the model predictions to the experimental data with a $\chi^2$-function
\begin{eqnarray}
\chi^2=\sum_i \frac{(\rho_i -\rho^0_i)^2}{\sigma_i^2} \; .
\end{eqnarray}
Here $\rho^0$ represents the data of the $i$th experimental
observable (taken from \cite{Tortola:2012te}), $\sigma_i$ the corresponding $1\sigma$ absolute error,
and $\rho_i$ the prediction of the model. In Fig.~\ref{fig:contour},
we present the allowed region of the mixing angles and the Dirac CP
phase at 1$\sigma$, 2$\sigma$, and 3$\sigma$ C.L., defined as the
contours in $\Delta \chi^2$ for two degrees of freedom with respect
to the $\chi^2$ minimum ($\chi^2_{\rm min} \simeq 1.7$).

We find that the best-fit value $\sin^2\theta_{12}=0.342$ slightly
deviates from its $1\sigma$ experimental interval, while
$\sin^2\theta_{23}=0.428$ agrees very well the current global fit
value. Note that the parameter space below the thin line in the left
plot of Fig.~\ref{fig:contour} is not allowed by the model itself,
due to the correlation between $\theta_{12}$ and
$\theta_{23}$. One reads from the right panel of
Fig.~\ref{fig:contour} that the Dirac CP phase is constrained to be
between $-0.5\pi$ and $0.5\pi$, while the best-fit values for
$\delta$ and $\sin^2\theta_{13}$ are around $\pm 0.27 \pi$ and
$0.024$. The contour is symmetric with respect to $\delta =0$.

\section{\label{sec:concl}Conclusions}
Non-zero $U_{e3}$ seems to make flavor symmetry models less economic, both in
size of the symmetry group as well as in particle content. We have presented here a
new method to accommodate non-zero $U_{e3}$ (and other deviations from tri-bimaximal mixing) that
keeps the minimality of typical models that were constructed to produce tri-bimaximal mixing.
Our idea takes into account that the product of two triplets contains
an anti-symmetric term, which vanishes for Majorana neutrinos. In case of Dirac neutrinos it remains,
and thus creates necessary deviations from tri-bimaximal mixing. This is not limited to the
particular mixing scheme (tri-bimaximal mixing) or the flavor group ($A_4$) that we used, or to the particular framework guaranteeing the
Dirac nature (a neutrinophilic 2 Higgs doublet model), but can be applied also to many other cases. Conceptually, our observation links the nature of the neutrino with the non-vanishing value
of $U_{e3}$.

\section*{Acknowledgments}
This work is supported by the Max Planck Society in the project
MANITOP through the Strategic Innovation Fund. We thank Christoph Luhn and Michael Schmidt
for helpful comments.

\appendix

\section{$A_4$ tensor products}
The basic tensor products of $A_4$, which we apply here, are given by~\cite{He:2006dk}
\begin{eqnarray}
{\bf 3} \times {\bf 3} & = & {\bf 3_{\rm a}} + {\bf 3_{\rm s}} +{\bf 1} + {\bf 1'} +{\bf 1''} \, , \nonumber \\
{\bf 1} \times {\bf 1} & = & {\bf 1} \;,\nonumber \\
{\bf 1'} \times {\bf 1'} & = & {\bf 1''} \; ,\\
{\bf 1''} \times {\bf 1''} & = & {\bf 1'} \;,\nonumber \\
{\bf 1'} \times {\bf 1''} & = & {\bf 1} \;,\nonumber
\end{eqnarray}
where (with $\omega=e^{i2\pi/3}$)
\begin{eqnarray}
({\bf 3} \times {\bf 3})_{3_{\rm s}} & = & \begin{pmatrix}x_2 y_3 +x_3 y_2, & x_3 y_1 +x_1 y_3, & x_1 y_2 +x_2 y_1 \end{pmatrix} , \nonumber \\
({\bf 3} \times {\bf 3})_{3_{\rm a}} & = & \begin{pmatrix}x_2 y_3 -x_3 y_2, & x_3 y_1
-x_1 y_3, & x_1 y_2 - x_2 y_1 \end{pmatrix} , \nonumber \\
({\bf 3} \times {\bf 3})_{1} & = & x_1 y_1+x_2 y_2+x_3 y_3 \, ,\\
({\bf 3} \times {\bf 3})_{1'} & = & x_1 y_1+ \omega x_2 y_2+ \omega^2 x_3 y_3 \, ,
\nonumber \\
({\bf 3} \times {\bf 3})_{1''} & = & x_1 y_1+ \omega^2 x_2 y_2+ \omega x_3 y_3
\; . \nonumber
\end{eqnarray}

\end{document}